# Application of quasi-optimal weights to searches of anomalies. Statistical criteria for step-like anomalies in cumulative spectra

A.V. Lokhov[a], F.V. Tkachov[b], P.S. Trukhanov[a]

[a] Department of Physics, Moscow State University, Moscow, 119991
[b] Institute for Nuclear Research RAS, Moscow, 117312

*Abstract.* The statistical method of quasi-optimal weights can be used to derive criteria for searches of anomalies. As an example we derive a convenient statistical criterion for step-like anomalies in cumulative spectra such as measured in the Troitsk-$\nu$-mass, Mainz and KATRIN experiments. It is almost as powerful as the locally most powerful one near the null hypothesis and appreciably excels the conventional $\chi^2$ and Kolmogorov-Smirnov tests. It is also compared with an ad hoc criterion of «pairwise correlations of neighbours»; the latter is seen to be less powerful if more sensitive to more general anomalies. As a realistic example, the criteria are applied to the Troitsk-$\nu$-mass data.

## 1. Introduction

Controversies around anomalies in experimental data — potential signals of new physics — are quite common in experimental physics. Given that a one-size-fits-all solution does not exist, the next best thing would be to have a more or less systematic approach to constructing correct statistical criteria for confirming/rejecting particular kinds of anomalies. In this paper we show how the foundational method of quasi-optimal weights [1] can be extended to a particular problem of this kind, namely, to the study of step-like anomalies in the Troitsk-$\nu$-mass experiment [2]-[4].

The first analysis [3], [4] of the Troitsk-$\nu$-mass data yielded a rather large and negative value for the neutrino mass squared, $m_\nu^2 \sim -(10 \div 20) \ eV^2$. This was interpreted as due to an excess of electrons near the end-point energy of the tritium $\beta$-decay spectrum; in cumulative spectra, such an excess takes the form of a step. Such a step is described by two parameters, height and position. Including these into the fit, a satisfactory value for the neutrino mass squared was obtained, $m_\nu^2 = -2.3 \pm 2.5_{fit} \pm 2.0_{syst} \ eV^2$ [2]-[4].

Recently a new data analysis has been completed using the method of quasi-optimal weights [1], adapted specifically for this task. This method (as opposed to the least-squares used in the first analysis) makes it possible to account for the non-Gaussian (Poisson) form of the experimental data distribution. Also the theoretical model of the experimental setup has been improved [5]. As a result of all the improvements, the new analysis yielded a physically relevant (within errors) value of the neutrino mass squared directly, i.e. without using the two additional parameters of the step.

Nevertheless the question of whether or not an anomalous contribution in the tritium $\beta$-decay spectrum is present in the Troitsk-$\nu$-mass experiment has not been definitely concluded as the $\chi^2$ criterion is not particularly sensitive to the presence of such contributions. It would be nice in such a situation (as in many other controversial cases in experimental physics) to settle the issue by applying a correct criterion specifically targeted to this particular anomaly.

For definiteness, we consider a simplified version of the problem that occurs in the real Troitsk-$\nu$-mass experiment (where the measured values have Poisson distribution etc.). This will allow us to perform explicit computations while avoiding nonessential complications.

However, the criterion that we are going to construct will be quite general and applicable, e.g., to situations with normally distributed data, etc.

Assume that some spectrum $\mu(E)$ is to be measured; the control variable $E$ will be called energy for convenience (in the Troitsk-$\nu$-mass experiment it is the retarding potential and $\mu$ is the cumulative spectrum that takes into account the geometrical resolution).

Suppose the measurements are done for some values $E_i$, $i=1,..,M$ ($M$ is the number of values $E_i$ in the given set).

Suppose that for each $E_i$ a number of events for a fixed period of time (the same for all $i$) is measured and this number is Poisson-distributed with $\mu_i = \mu(E_i)$.

The anomalous contributions for which we need a special criterion have the form of a step (in the Troitsk-$\nu$-mass cumulative spectrum the step is a result of integration of a $\delta$-shaped anomaly):

$$\mu'(E) = \mu(E) + \Delta \cdot \theta(E_{st} - E), \qquad (1)$$

where $\mu(E)$ is the spectrum without the anomalous contribution, $\theta(x)$ is the Heaviside function, $E_{st}$ is the step position and $\Delta$ is its height. In terms of the set $E_i$, define $m$ according to $E_m$, $E_m \leq E_{st} < E_{m+1}$ (Fig.1).

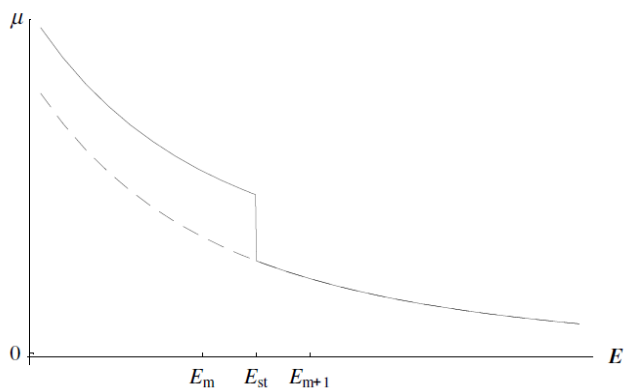

**Figure 1.** The typical form of the end of the cumulative spectrum of tritium $\beta$-decay with (solid line) and without (dashed line) step-like anomalous contribution.

We assume that the data are fitted without account of the anomaly. The goal is then to find a powerful, general and handy test which would be sensitive to anomalous contributions of the described type. To compare the criteria, the so-called power functions will be used.

## 2. The conventional criteria

There are conventional, nonspecific statistical criteria, such as $\chi^2$ and Kolmogorov-Smirnov tests [6].

The $\chi^2$ statistics is

$$\chi^2 = \frac{1}{M-1} \sum_i \xi_i^2, \text{ where } \xi_i = (N_i - \mu_i)/\sqrt{\mu_i} \qquad (2)$$

The Kolmogorov test can not be directly applied in our problem because it requires a sample of one continuous variate. In our case there is a set of different and discrete variates (obeying Poisson distributions with different means).

However, we can use Kolmogorov's idea as follows. At first we switch over from variates $N_i$ to normalized quantities $\xi_i = (N_i - \mu_i)/\sqrt{\mu_i}$. Then, similarly to the Kolmogorov-Smirnov criterion, define:

$$D = \max_i \left| F_0(\xi_i) - S_i \right|, \qquad (3)$$

where $F_0(\xi)$ is the theoretical cumulative distribution function for $\xi$, and $S_i$ is the value of the sample distribution function (for the sample $\{\xi_i\}$) at $\xi_i$. The criterion based on statistics $D$ will be called the modified Kolmogorov-Smirnov test.

For both statistics — $\chi^2$ and $D$ — one can conduct the standard test of hypotheses:

H0: the height of the step is $\Delta = 0$

H1: the height of the step is $\Delta > 0$.

It is obvious, however, that such criteria, while being rather general, are not specifically tuned to searches of step-like anomalies.

## 3. Locally most powerful (LMP) criterion for step search

We are going to construct a criterion along the lines of the method of quasi-optimal weights. In the language of mathematical statistics, such criteria are locally most powerful [7] («locally» here means «near the null-hypothesis»; we will use the abbreviation LMP in what follows).

At first assume that the step position is known (for example from the first data analysis in the Troitsk-$\nu$-mass case). It is described by $m$ defined after eq.(1). We take the following parameterization of the distributions of the number of events:

$$f_i(N) = \frac{\mu_i'^N e^{-\mu_i'}}{N!}, \text{ where} \qquad (4)$$

$$\mu_i' \to \begin{cases} \mu_i + \Delta_-, & i > m \\ \mu_i + \Delta_+, & i \leq m \end{cases} \quad (5)$$

The parameterization corresponds to a situation where the data are first fitted within the null hypothesis, i.e. without regard of the step ($\mu_i$) and then the step height $\Delta = \Delta_+ - \Delta_-$ is fitted separately.

The method of quasi-optimal weights [1] immediately yields optimal weights to estimate the parameters $\Delta$:

$$\omega_i^+(N) = \frac{\partial \ln f_i}{\partial \Delta_+} = \begin{cases} 0, & i > m \\ \frac{N}{(\mu_i + \Delta_+)} - 1, & i \leq m \end{cases} \quad (6)$$

$$\omega_i^-(N) = \frac{\partial \ln f_i}{\partial \Delta_-} = \begin{cases} 0, & i \leq m \\ \frac{N}{(\mu_i + \Delta_-)} - 1, & i > m \end{cases} \quad (7)$$

The theoretical means of such weights are always equal to *0* [1].
The corresponding experimental means of the weights are the following:

$$h_-^{\exp} = \frac{1}{M} \sum_{i=1}^{M} \omega_i^- = \frac{1}{M} \sum_{i=m+1}^{M} \left( \frac{N_i}{\mu_i + \Delta_-} - 1 \right), \quad (8)$$

$$h_+^{\exp} = \frac{1}{M} \sum_{i=1}^{M} \omega_i^+ = \frac{1}{M} \sum_{i=1}^{m} \left( \frac{N_i}{\mu_i + \Delta_+} - 1 \right). \quad (9)$$

Equating $h^{\exp}$ to *0* one obtains two equations for $\Delta_+$ and $\Delta_-$:

$$\begin{cases} \sum_{i=m+1}^{M} \left( \frac{N_i}{\mu_i + \Delta_-} - 1 \right) = 0 \\ \sum_{i=1}^{m} \left( \frac{N_i}{\mu_i + \Delta_+} - 1 \right) = 0 \end{cases} \quad (10)$$

Solving the equations relative to $\Delta_+$ and $\Delta_-$ one obtains a quasi-optimal estimation of the step height for the given sample of data $\{N_i\}$ as $\bar{\Delta} = \Delta_+ - \Delta_-$. $\bar{\Delta}$ is the LMP criterion for the step.

The LMP criterion as defined above has two drawbacks. The first, minor drawback is that it is not defined by explicit formulae. This can be remedied by assuming smallness of $\Delta_+$ and $\Delta_-$ relative to $\mu_i$ and performing expansion:

$$\begin{cases} \sum_{i=m+1}^{M}\left(\dfrac{N_i}{\mu_i}-\dfrac{N_i}{\mu_i^2}\Delta_- -1\right)=0 \\ \sum_{i=1}^{m}\left(\dfrac{N_i}{\mu_i}-\dfrac{N_i}{\mu_i^2}\Delta_+ -1\right)=0 \end{cases} \rightarrow \begin{cases} \Delta_- = \left(\sum_{i=m+1}^{M}\dfrac{N_i}{\mu_i}-(M-m)\right)\Big/\sum_{i=m+1}^{M}\left(\dfrac{N_i}{\mu_i^2}\right) \\ \Delta_+ = \left(\sum_{i=1}^{m}\dfrac{N_i}{\mu_i}-m\right)\Big/\sum_{i=1}^{m}\left(\dfrac{N_i}{\mu_i^2}\right) \end{cases} \quad (11)$$

The second, much more serious drawback is that it is too sensitive to the step position; this makes it inconvenient in practice. For instance, in the Troitsk-$\nu$-mass experiment the step position was seen to vary in time. For practical applications it may be desirable to reduce such a sensitivity even at the price of some loss of efficiency.

Despite its drawbacks, the LMP criterion is a convenient starting point for the construction of an efficient «quasi-optimal» criterion with a reduced sensitivity to the step position.

## 4. «Quasi-optimal» criterion

In the spirit of the method of quasi-optimal weights [1], let us simplify the LMP criterion obtained above, preserving its characteristic form.

The LMP test (11) can be represented in terms of $\xi_i = (N_i - \mu_i)/\sqrt{\mu_i}$ as a weighted sum of the form $\sum_i w_i \cdot \xi_i$ (an overall constant addendum plays no role). The most salient feature of the weights $w_i$ is that they change sign at the step position. A natural idea is to adjust the weights so that they become a piecewise linear function of the point number:

$$S_{q-opt} = \sum_{i=1}^{M} w_i \cdot \xi_i \quad (12)$$

$$w_i = \begin{cases} (m-i)/m, & i \leq m, \\ (m-i)/(M-m), & i > m, \end{cases} \quad (13)$$

with a break at some point $E_m$ in the hypothetical domain of variation of the step position $E_{st}$ (this domain is known e.g. in the case of the Troitsk-$\nu$-mass experiment).

The weights thus modified still change the sign at the step position, but the dependence of the sum on the step position is reduced because the contributions from its neighbourhood are suppressed.

In what follows we will simply call $S_{q-opt}$ the *quasi-optimal criterion*.

## 5. Alternative criterion (pairwise correlations of neighbours)

For comparison we present another test constructed somewhat speculatively. It will be seen to be less powerful for the search of step-like anomalies, but it is sensitive to more general anomalous contributions. Let the anomaly be a deviation of several neighbour points to one side of the fitting curve. Then consider the test with the following statistics:

$$S_{pair} = \sum_i \xi_i \cdot \xi_{i+1} \qquad (14)$$

A deviation of a pair of neighbour points $(\xi_i, \xi_{i+1})$ to the same side of the fitting curve makes a positive contribution to $S_{pair}$. Therefore $S_{pair}$ will be sensitive to anomalies where groups of neighbour data points deviate from the fitting curve in the same direction; a step anomaly is a special case of this kind of anomalies.

## 6. Comparing the criteria

The various criteria described above can be systematically compared using the standard statistical instrument of the so-called power functions.

The power function of a criterion describes its efficiency in separating the null-hypothesis (the height of the step is *0*) from the alternative H1 (the height of the step is $\Delta > 0$). The value of the power function at $\Delta > 0$ is defined as the probability to reject the null-hypothesis when the alternative H1 is correct.

We have constructed the power functions for all our tests using Monte Carlo modelling.

For a realistic example, we used the preliminary fit values of $\mu_i$ from one of the runs of the Troitsk-$\nu$-mass experiment as theoretical parameters of Poisson distributions. The data consist of sets of 44 triplets $\mu_i$, $N_i$ and the corresponding energies $E_i$ ( $E_i$ fall into the range $18400\ eV \leq E_i \leq 18770\ eV$ ).

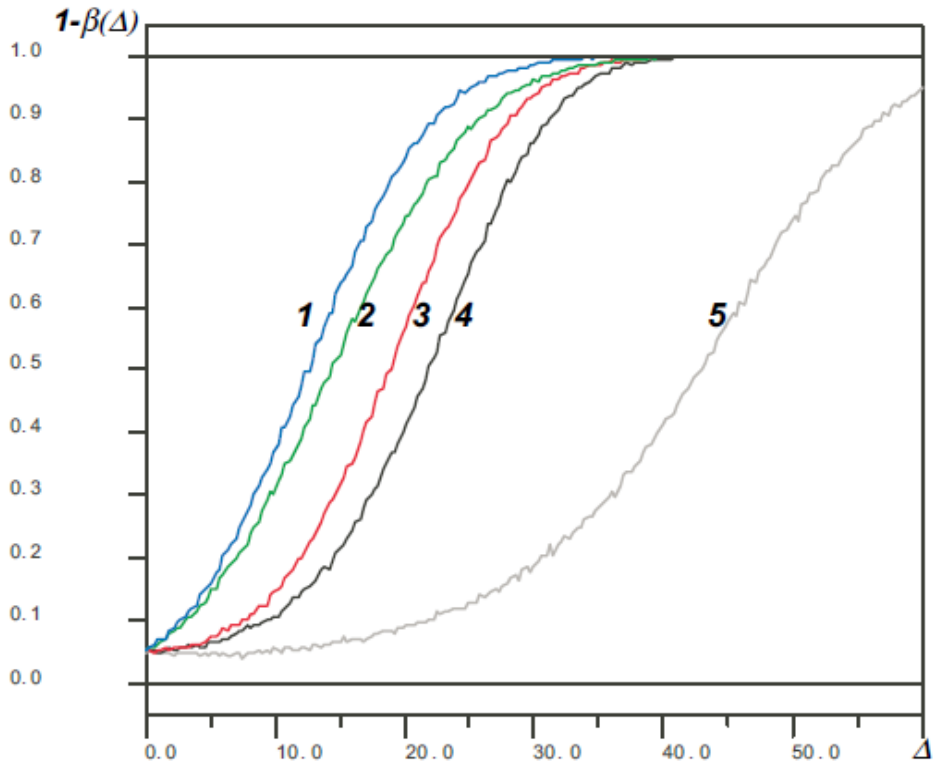

**Figure 2.** The power functions for the LMP criterion (blue, 1), the «quasi-optimal» criterion (green, 2), the criterion of «pairwise neighbours correlations» (red, 3), the $\chi^2$ (black, 4), and the modified Kolmogorov-Smirnov criterion (grey, 5). $\Delta$ is the height of the step; see Eq.(1).

The power functions of the five criteria allow one to compare their efficiency graphically (Fig.2). The LMP criterion (1) appears to be the best among all considered, the «quasi-optimal» one (2) is slightly less powerful, and the criterion of «pairwise neighbours correlations» (3) comes in third. The least powerful are the criteria which exploit no information about the anomaly: the $\chi^2$ (4) and the modified Kolmogorov-Smirnov (5) tests.

**Sensitivity to the step position.** The stability of the five criteria with respect to shifts of the step position (when the assumed position $E_m$ differs from the unknown true value $E_{st}$) can also be investigated with the help of the power functions. To this end we construct (using Monte Carlo again) the power functions for two values of the shift.

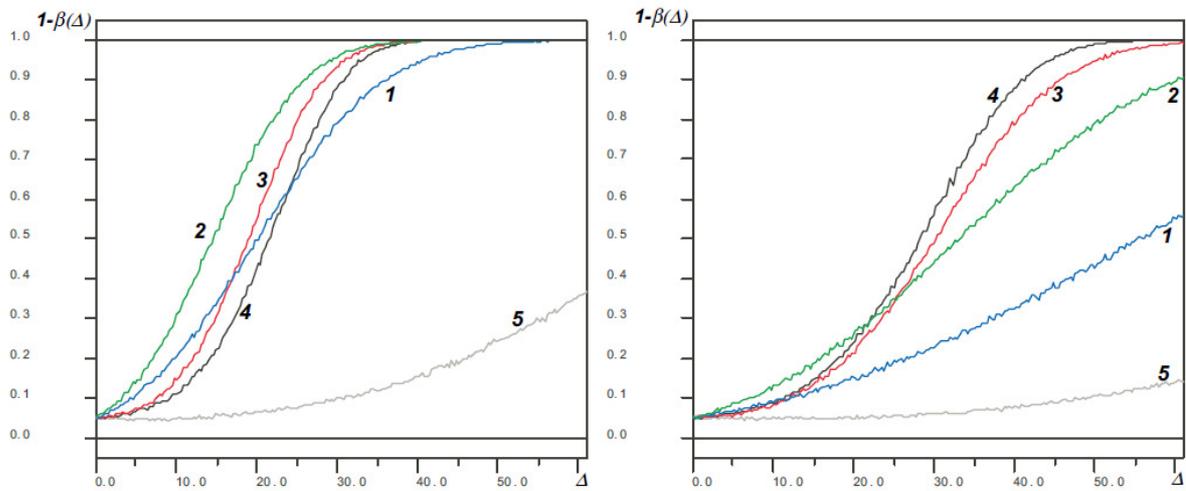

**Figure 3.** The power functions of the five criteria for the cases when the actual step position is shifted from the assumed position $E_m$ by 12 $eV$ (left) and 25 $eV$ (right).

From Fig. 3 it is seen that the LMP criterion (the most powerful one if the step position is known exactly) loses its advantage with the shift of the step. At the same time the «quasi-optimal» test remains sensitive to the anomalous contribution (Fig.3 left, the curve 2 (green)).

With the significant shift of the position (25 $eV$), the criteria independent from the anomaly parameters become more powerful (Fig.3 right, the red (3) and black (4) curves).

## 7. Applications to data

In a realistic experimental situation where a spectrum of the described type is measured, one chooses a test, say, $S_{q-opt}$ and constructs the corresponding distribution function $F_1(S_{q-opt})$ via Monte Carlo modelling. This is done by repeatedly simulating the measured data set (at the null hypothesis), evaluating $S_{q-opt}$ for each simulation, and then building e.g. a histogram to obtain the distribution function $F_1(S_{q-opt})$. Then, given a real measured data set, one computes $S_{q-opt}$ for it and then finds $\alpha^q_{q-opt} = F_1(S^q_{q-opt})$. Since this $\alpha$ is the probability for the values of the test not to exceed the corresponding $S$ (at the null hypothesis), a presence of anomaly in the data would be directly signalled by $\alpha$ exceeding the confidence level chosen (e.g. 0.95).

One can of course repeat this procedure with any number of tests. Also, in a multi-run experiment, each run may have to be handled separately. Then each test + run combination would be an independent instance of application of the described procedure. For instance, the new data analysis for the Troitsk-$\nu$-mass data selected eleven runs [5], and each run's data set can be treated with the quasi-optimal or «pairwise neighbours correlations» criteria, yielding 22 tests in total. The table below lists all 22 numbers for example's sake, but note that the fact that for each criterion not one but eleven independent numbers are available invites for an additional statistical study which, however, would take us too far from the subject of the present paper; see ref. [8].

| Run# | 22 | 23 | 24-1 | 24-2 | 25 | 28 | 29 | 30 | 31 | 33 | 36 |
|---|---|---|---|---|---|---|---|---|---|---|---|
| $\alpha^q_{q-opt}$ | 0.382 | 0.359 | 0.381 | 0.522 | 0.478 | 0.266 | 0.510 | 0.371 | 0.365 | 0.570 | 0.207 |
| $\alpha^q_{pair}$ | 0.883 | 0.571 | 0.471 | 0.604 | 0.920 | 0.829 | 0.141 | 0.994 | 0.113 | 0.702 | 0.810 |

**Table 1.** The values $\alpha^q_{q-opt} = F_1(S^q_{q-opt})$ and $\alpha^q_{pair} = F_2(S^q_{pair})$ for the 11 runs of the Troitsk-$\nu$-mass experiment. The value of $\alpha$ exceeding a chosen confidence level (e.g. 0.95) directly signals a presence of anomaly. With such an array of numbers for each test, however, a further statistical analysis may be worthwhile (cf. ref. [8]).

## 8 Conclusions

We have shown how the method of quasi-optimal weights can be used to construct special-purpose statistical criteria, namely, the criteria for determining presence of step-like anomalies in cumulative spectra. We have constructed:

(1) the theoretically best but somewhat cumbersome LMP criterion;

(2) a «quasi-optimal» test obtained by a judicious simplification of the LMP test;

(3) a «pairwise neighbours correlations» test constructed somewhat speculatively.

We compared them with the standard $\chi^2$ and Kolmogorov-Smirnov tests using the standard tool of the power functions:

— The «quasi-optimal» test is only slightly less efficient then the LMP criterion, while being much simpler to compute, and it is also more stable with respects to shifts of the unknown step position.

— The «pairwise neighbours correlations» test, while being less efficient then the «quasi-optimal» test, is suitable for anomalous contributions of a more general shape.

— The $\chi^2$ test can be of interest only in the cases of very large shifts of the actual position of the anomaly from the one assumed in the «quasi-optimal» test.

— The modified Kolmogorov-Smirnov criterion appears to be the least sensitive in the entire range of parameters.

To provide realistic examples, the constructed criteria have been applied to data sets borrowed from the new data analysis of the Troitsk-$\nu$-mass experiment [5]. We refrain, however, from making a physical conclusion concerning presence or absence of step-like anomalies in that experiment since that would require an additional statistical analysis; this is done in a separate paper [8].

## Aknowledgements

The authors thank A.A.Nozik for the fitted experimental data samples used in the examples, V.S.Pantuev for critical remarks, and the collaborators of the Troitsk-$\nu$-mass experiment and members of the Department of Theoretical Physics of INR RAS for discussions.